\documentclass[conference,letterpaper,onecolumn]{IEEEtran}
\usepackage{cite,amsmath,amsthm,amssymb,amsfonts,url,graphicx,overpic}

\usepackage[utf8]{inputenc} 
\usepackage[T1]{fontenc}
\usepackage{url}
\usepackage{ifthen}
\usepackage{parskip}
\usepackage{booktabs} 

\graphicspath{{./figures/}}

\newcommand{\tJ}{\tilde{J}}

\newtheorem{theorem}{Theorem}[section]

\newtheorem{definition}[theorem]{Definition}

\newtheorem{remark}[theorem]{Remark}

\newcommand{\parenth}[1] {\left(#1\right)}

\newcommand{\brackets}[1] {\left[#1\right]}

\newcommand{\reals}{\mathbb{R}}

\newcommand{\half} {\frac{1}{2}}

\newcommand{\bitm}{\begin{itemize}}
\newcommand{\eitm}{\end{itemize}}
\newcommand{\benum}{\begin{enumerate}}
\newcommand{\eenum}{\end{enumerate}}
\newcommand{\beqa}{\begin{eqnarray}}
\newcommand{\eeqa}{\end{eqnarray}}
\newcommand{\beqas}{\begin{eqnarray*}}
\newcommand{\eeqas}{\end{eqnarray*}}
\newcommand{\baln}{\begin{align}}
\newcommand{\ealn}{\end{align}}
\newcommand{\balns}{\begin{align*}}
\newcommand{\ealns}{\end{align*}}

\title{A Kuramoto–von Mises Time Series Model for Probabilistic Modeling of Coupled Oscillators}

\author{%
  \IEEEauthorblockN{Yun Hwang}
  \IEEEauthorblockA{Stanford University \\
                    Email: yunhwang@stanford.edu}
  
  \and
  \IEEEauthorblockN{Todd P.~Coleman}
  \IEEEauthorblockA{Stanford University \\
                    Email: toddcol@stanford.edu}
}

\begin{document}
\maketitle 

\begin{abstract}
A system of coupled oscillators provides a fundamental framework for modeling a wide range of physical and biological phenomena. In neuroscience, the central nervous system exhibits synchronized oscillatory activity with adjacent brain regions, giving rise to traveling wave dynamics for instance during sleep. Similarly, in the gastrointestinal system, neuromuscular cells coordinate their oscillations to generate propagating waves of slow wave activity.  To estimate probability distributions of multivariate phase relationships, existing approaches typically rely on equilibrium thermodynamics, expressing the system in a Boltzmann form through a pairwise exponential family distribution. However, these assumptions are often violated in real-world systems, which are inherently dynamic and frequently transition between equilibrium and non-equilibrium regimes. To address this, we propose an efficient method for estimating the probability distribution of coupled oscillators that does not assume thermodynamic equilibrium. Using a Langevin dynamics–based construction, the approach enables accurate modeling even in non-equilibrium regimes. The maximum likelihood estimation method is shown to have a closed form algebraic solution in the high sampling rate regime, a condition commonly satisfied by modern data acquisition systems, which makes it readily applicable in practice. We demonstrate its robustness on simulated data, where it outperforms existing approaches in non-equilibrium settings, and further illustrate its utility for characterizing dynamic brain traveling waves in response to brain stimulation and in hypothesis testing within the context of electrophysiologic recordings of the human stomach.
\end{abstract}

\section{Introduction}
Numerous physical and biological phenomena can be modeled as a system of coupled oscillators. For instance, the conformational geometry of proteins is governed by dynamically evolving angles along the amino acid backbone and side chains. These angular degrees of freedom can be naturally represented as phases, whose interactions critically influence protein folding processes \cite{Tsuboyama2023,kim2015multivariate}. In neuroscience, coordinated neural activity often manifests as propagating waves across spatially distributed brain regions. As an example, the human brain repeatedly coordinates sequential activation of neighboring regions, forming rotating waves when sleep spindles occur. \cite{Muller2016}. Wave phenomena in the brain are known to also play a role in information integration over space and time in the somatosensory cortex and middle temporal visual area \cite{gonzales2025touch,davis2024horizontal}. These functions of traveling wave in spatiotemporal integration have also been demonstrated in deep neural network models \cite{keller2024waves, jacobs2025traveling}. In the gastrointestinal system, networks of interstitial cells of Cajal (ICC) exhibit intrinsic oscillatory behavior, with individual cells generating slow-wave membrane potential oscillations at approximately 0.05 Hz \cite{gharibans2019spatial}. These cells are electrically coupled to one another, as well as to adjacent smooth muscle cells, via gap junctions, forming a functional syncytium. The resulting collective dynamics can be understood as a network of coupled oscillators, where phase interactions govern the emergence of coherent rhythmic activity \cite{parsons2016spatial,parsons2018phase}.

The Kuramoto model provides a principled framework for analyzing a system of coupled oscillators, and a growing body of work has focused on estimating the probability distribution of multivariate phase relationships \cite{kuramoto1975self, cadieu2010phase, perley_coleman_2025_phase}. Early approaches leveraged score matching method to infer coupling structure without requiring explicit normalization of the likelihood \cite{cadieu2010phase}. However, this work doesoes not have theoretical guaranties on the performance or sample complexity. More recent work  leverages interaction screening objectives and tree-structured approximations to estimate distributions over multivariate phase relationships \cite{perley_coleman_2025_phase}. This approach generalizes recent work on the estimation method for the connectivity structure in Ising models as a convex optimization problem \cite{vuffray2016interaction}. Their approach provides useful theoretical properties in selecting the strength of  a $\ell_1$ regularizer to have a sparse estimate of the underlying connectivity structure in a system of coupled oscillators. It eliminates the need for a train–test split to assess the generalizability of the learned model with the selected regularization strength.

However, these methods rely on two critical assumptions: (i) that the coupling matrix is symmetric, and (ii) that the system has reached equilibrium, allowing the distribution to be expressed in Boltzmann form. These assumptions are often violated in real-world systems as many biological networks exhibit inherently asymmetric interactions—for example, excitatory and inhibitory synaptic connections in neural circuits—and are frequently driven by external inputs, placing them in non-equilibrium regimes \cite{schmitz2022development, aggarwal2022visual}. As a result, equilibrium-based formulations may provide a limited or biased characterization of the underlying dynamics of a system of coupled oscillator.

In this work, we propose a method for estimating underlying phase distributions without assuming symmetric connectivity or an equilibrium state, instead modeling the dynamics using a more general Langevin framework. The maximum likelihood estimation method is shown to have a closed form algebraic solution in the high sampling rate regime, a condition commonly satisfied by modern data acquisition systems, which makes it readily applicable in practice. Our linear method is efficient and generalizable to both equilibrium and nonequilibrium conditions. In this work, we first review the relevant background and related work in section 2. We then derive the proposed estimation procedure in sections 3 and 4. Finally, we evaluate the performance of the method by estimating the underlying connectivity structure in high-dimensional systems of coupled oscillators in Section 5. 

\vspace{0.5em}

\section{Preliminaries on Kuramoto Models, Boltzmann distribution, and von Mises Distribution}
\subsection{Kuramoto Model for a System of Coupled Oscillators}
The Kuramoto model \cite{kuramoto1975self} is one of the most widely used models of interacting elements with nonlinear dynamics, specifically within the context of synchronization of coupled oscillators \cite{acebron2005kuramoto}.  It has been utilized in numerous physical, biological, chemical, and societal systems, where there does not exist a central coordinating mechanism and each interacting element can be represented as a phase oscillator \cite{strogatz2001exploring}. Suppose there are $N$ oscillators under the standard Kuramoto model. The $i$th oscillator has a phase, which at time $t$ is given by $y_{i,t}$, measured in radians.  Oscillator $i$ has an intrinsic oscillation angular frequency of $\omega_i$.  In addition, it is coupled to other oscillators with sine attractive coupling:
\begin{eqnarray}
\dot{y}_i = \omega_i + \sum_{j=1}^N \kappa_{ij} \sin(y_j-y_i) \label{eqn:original:Kuramoto}
\end{eqnarray}

\subsection{Control Theory Interpretation of Kuramoto Model}

Suppose that $y_j(t-\Delta)$ and $y_i(t-\Delta)$ are near each other with $y_j(t) > y_i(t)$  .  We can interpret this model from a control perspective where the term $\sin(y_j-y_i)$ operates as a control signal with the purpose to keep oscillators $i$ and $j$ synchronized if $\kappa_{i,j} > 0$. As an illustration, oscillator $i$ needs to oscillate faster in order to synchronize with oscillator j. The small, positive $\sin(y_j(t)-y_i(t))$ term will provide extra angular velocity to facilitate the synchronization:

\begin{eqnarray}
  \dot{y}_i &=& \omega_i + \frac{1}{N} \sum_{j=1}^N \kappa_{ij} \underbrace{\sin(y_j-y_i)}_{\shortstack{a small control signal\\telling oscillator $i$ to speed up}}
\end{eqnarray}

\subsection{Kuramoto Model with Temporal Delay}

\newcommand{\tauij}{\tau_{ij}}

Our current Kuramoto formulation assumes instantaneous influence from oscillator $j$ to oscillator $i$. However, in physiological systems, interactions are often subject to non-negligible temporal delays in the effect of oscillator $j$ on oscillator $i$ \cite{zhang2017ionic}. In this work, we consider a system with delays in coupled interaction as described in \cite{budzinski2023analytical}, where there's a temporal delay from oscillator $j$ to oscillator $i$ given by $\tauij$. The Kuramoto system with temporal delays can be written as
\begin{eqnarray}
\dot{y}_i = \omega_i + \sum_{j=1}^N
\kappa_{ij} \sin\parenth{ y_j(t-\tauij)-y_i(t)}
\end{eqnarray}

Since the $j$th oscillator on average oscillates at frequency $\omega_j$, the inner term in the $\sin$ function can be approximated as follows \cite{budzinski2023analytical}:
\[ y_j(t-\tauij) \simeq y_j(t)-\omega_j \tauij \]
As such, we now have
\begin{eqnarray}
\dot{y}_i &=& \omega_i + \sum_{j=1}^N \kappa_{ij} \sin\parenth{ y_j(t)-\omega_j \tauij-y_i(t)} \nonumber\\ 
&=& \omega_i + \sum_{j=1}^N
\kappa_{ij} \sin\parenth{ y_j(t)-y_i(t)-\omega_j \tauij} \nonumber \\
&=& \omega_i + \sum_{j=1}^N
\kappa_{ij} \sin\parenth{ y_j(t)-y_i(t)-\mu_{ij}}
\label{eqn:kur_delay_mod}
\end{eqnarray}
where $\mu_{ij} = \omega_j\tau_{ij}.$
Note that \eqref{eqn:kur_delay_mod} is equivalent to the formulation introduced from earlier work \cite{cadieu2010phase}. For later convenience, we further express this using the following trigonometric identity:
\[ \sin(a-b) = \sin a \cos b - \cos a \sin b.\]
With
$a= y_j(t)-y_i(t)$ and $b=\mu_{ij}$, we have
\[ \sin\parenth{ y_j(t)-y_i(t)-\mu_{ij}}
= \sin \parenth{y_j(t)-y_i(t)}
\cos \parenth{\mu_{ij}}
- \cos \parenth{y_j(t)-y_i(t)}
\sin \parenth{\mu_{ij}}.
\]
Instead of directly estimating $\kappa_{ij}$ and $\mu_{ij}$, we can go to angular coordinates and write $\kappa^s_{i,j}$ $\kappa^c_{i,j}$ as follows:
\beqas
\kappa^s_{ij} &=& -\kappa_{ij} \sin \parenth{\mu_{ij}} \\
\kappa^c_{ij} &=& \kappa_{ij} \cos \parenth{\mu_{ij}}
\eeqas 
Now we can rewrite it as:
\begin{eqnarray}\label{eqn:kur_td}
\dot{y}_i 
= \omega_i + \sum_{j=1}^N
\kappa^c_{ij} \sin\parenth{ y_j(t)-y_i(t)}
+ \kappa^s_{ij} \cos\parenth{ y_j(t)-y_i(t)}
\end{eqnarray}


\subsection{Langevin and Boltzmann Formulation} 
As previously mentioned, prior work estimate the Kuramoto system by assuming they are in equilibrium in a thermodynamic sense. We briefly mention concepts of Langevin dynamics and Boltzmann distribution.

\begin{definition}[Langevin dynamics] \label{def:langevin}
General Langevin dynamics have the form:
\[
d\underline{Y}_t = b(\underline{Y}_t)\,dt + \sigma\,d\underline{W}_t,
\]
with $b(\underline{y})$ and $\underline{W}_t $ representing deterministic flow and Wiener processes, respectively. 
\end{definition}

\begin{remark}[Gradient field] \label{def:grad_field}
Let $\phi: \mathbb{R}^n \to \mathbb{R}$ be a scalar function. A gradient field is a vector field of the form $F = \nabla \phi$. One of the main properties of a gradient field is that its curl is always zero: $\nabla \times F = 0$.
\end{remark}

\begin{definition}[Boltzmann distribution]\label{boltzmann}
If $b(\underline{y})$ is a negative gradient field, i.e.,
\[
dY_t = -\nabla U(\underline{Y}_t)\,dt + + \sigma\,dW_t,
\]
$f(\underline{y}_t)$ converges to an equilibrium with phase distribution:
\[
f(\underline{y}) \propto \exp\!\left(-\frac{U(\underline{y})}{D}\right)
\]
and we call this Boltzmann distribution. 
\end{definition}
By letting $b(\underline{y}_t \, ; K) = \sum_{i,j} \kappa_{i,j}\sin(y_{j,t}-y_{i,t} - \mu_{ij})$, one can estimate the distribution in equilibrium of the form:
$$f(\underline{y}\,;K) \propto \exp\parenth{-\frac{1}{D}\sum_{i,j} \kappa_{ij} \cos(y_i-y_j-\mu_{ij})}$$
which is the distribution that previous approaches focus on \cite{cadieu2010phase,perley_coleman_2025_phase}. However, note that estimating the Kuramoto system with this distribution relies on two key assumptions. First assumption is that $K$ has to be a symmetric matrix (i.e., $\kappa_{ij} = \kappa_{ji}$ ) in order for $b(\underline{y})$ to be a negative gradient field. This notion comes from the fact that the curl of $b(\underline{y})$ cannot have a rotational component (i.e., $\nabla \times b(\underline{y})=0$) to become a gradient field as mentioned in \ref{def:grad_field}. Second assumption is that the system of interest is at an equilibrium, which may not be the case in many real-world physical phenomena. In this work, we propose a method that does not make these assumptions, providing us with a more general method to estimate the underlying phase dynamics.

\subsection{From Deterministic to Stochastic Kuramoto Modeling}
\eqref{eqn:kur_delay_mod} is a deterministic differential equation.  
A natural probabilistic formulation is to model the continuous-time evolution using the Langevin formulation as in \ref{def:langevin} \cite{oksendal2013stochastic}:
\begin{eqnarray*}
dY_{i,t} = \parenth{\omega_i + \sum_{j=1}^N \kappa_{i,j} \sin(Y_{j,t}-Y_{i,t}-\mu_{ij})}dt + \sigma dW_{i,t}
\end{eqnarray*}
where $W_1,\ldots,W_N$ are independent Wiener processes and each $\sigma$ is a constant.  Note then that the above equation is an instance of the standard stochastic differential equation (SDE).  Since Brownian motion is not circular, we can assume we have unwrapped the phase or equivalently we represent the right hand side of the equation mod $[0,2\pi]$.

What does the above SDE mean? Let $N=2$, let $g: \mathbb{R}^2 \to \mathbb{R}^2$ be a vector valued map and consider this general SDE
\begin{equation} 
d \underline{Y}_t = g(\underline{Y}_t)dt + \sigma d\underline{W}_t \label{eqn:standard:SDE}
\end{equation}
where $\underline{W}$ is a vector-valued Wiener process. Discretizing time into intervals of length $\Delta$, we have

\begin{equation}
\label{eqn:kuramoto_noise}
\underline{Y}_{t+\Delta} = \underline{Y}_t + \Delta g(\underline{Y}_t) + \sigma \left(\underline{W}_{t+\Delta} - \underline{W}_t\right).
\end{equation}

If $g=(g_1,g_2)$ where $g_1: \reals^2 \to \reals$ and $g_2: \reals^2 \to \reals$, we have:
\begin{subequations}\label{eqn:vectorvaluedSDE}
\begin{eqnarray} 
Y_{1,t+\Delta} &=& Y_{1,t} + \Delta g_1(\underline{Y}_t) +\sigma \left(W_{1,t+\Delta} - W_{1,t} \right) \\
Y_{2,t+\Delta} &=& Y_{2,t} + \Delta g_2(\underline{Y}_t) + \sigma \left(W_{2,t+\Delta} - W_{2,t} \right) 
\end{eqnarray}
\end{subequations}
where $\left(W_{1,t+\Delta} - W_{1,t} \right)$ and $\left(W_{2,t+\Delta} - W_{2,t} \right)$ are statistically independent and both with $\mathcal{N}(0,\Delta)$ statistics.
This statistical conditional independence will be leveraged later so we anchor a remark here:
\begin{remark}
\label{remark:SDE:conditional-indepdnenece}
In \eqref{eqn:vectorvaluedSDE}, $Y_{1,t+\Delta}$ and $Y_{2,t+\Delta}$ are statistically independent given $\underline{Y}_{[0:t]}$.
\end{remark}

Applying this logic to the stochastic Kuramoto model, we have to think about performing a modulo operation on the right hand side of \eqref{eqn:vectorvaluedSDE} due to the $\left(W_{i,t+\Delta} - W_{i,t} \right)$ term which has a Gaussian distribution.

\subsection{von Mises Distribution to model phases} 
If we now consider the Kuramoto analogue of \eqref{eqn:vectorvaluedSDE},  notice for oscillator $i$ we will have
\beqa
Y_{i,t+\Delta} &=& Y_{i,t} + \Delta \parenth{\omega_i + \sum_{j=1}^N \kappa_{ij} \sin(Y_{j,t}-Y_{i,t}-\mu_{ij})} +\sigma \left(W_{i,t+\Delta} - W_{i,t} \right) \mod [0,2\pi] 
\eeqa
Notice that this can be interpreted as a time series model with intrinsic and ensemble network covariates analogous to conditional intensity point process models \cite{truccolo2005point} in that when predicting $\underline{Y}$ at time $t+\Delta$, the right hand side contains regressors from the recent past of $\underline{Y}_{[0:t]}$.  Specifically, note that given $\underline{Y}_{[0:t]}=\underline{y}_{[0:t]}$, the random variable $Y_{i,t+\Delta}$ on average is $y_{i,t} + \Delta \parenth{\omega_i + \sum_{j=1}^N \kappa_{i,j} \sin(y_{j,t}-y_{i,t}-\mu_{ij})}$ and its variability  comes from $\sigma \left(W_{i,t+\Delta} - W_{i,t} \right)$. If we were on the real line, this conditional distribution would be a Gaussian centered at the average and with variance $\Delta \sigma^2$.  In our case, phase is in between $[0,2\pi]$.  It is well known that the  wrapped Gaussian distribution (e.g. the Gaussian $\mod [0, 2\pi]$) is well approximated by the von Mises distribution. 
\begin{definition}[von Mises distribution]
A random variable $y$ taking values on the circle $[0, 2\pi)$ is said to follow a von Mises distribution with mean direction $\mu \in [0, 2\pi)$ and concentration parameter $\kappa \ge 0$, 
if its probability density function is given by
\[
f(\theta \mid \mu, \nu)
= \frac{1}{2\pi I_0(\nu)} \exp\!\big(\kappa \cos(\theta - \mu)\big),
\quad y \in [0, 2\pi),
\]
where $I_0(\nu)$ is the modified Bessel function of the first kind of order zero.
\end{definition}

Just like the Gaussian distribution, the von Mises density has a single maximum which is located at is expectation $\mu$, and its distribution has another parameter $\kappa$ which is the inverse of the variance.   In the above equation, $I_0$ is the Bessel function of order $0$ (which does not depend on $\mu$).
The approximation of wrapped Gaussians as von Mises distributions in the limit of low variance can be explicitly stated in how a wrapped Gaussian with mean $\mu$ and variance $\Delta$ has a density that uniformly approaches the density of a von Mises mean $\mu$ and inverse variance $\frac{1}{\tau}$ as $\tau \to 0$.  Specifically, for any $\mu$, we have that
$$ \lim_{\Delta \to 0} \sup_{y \in [0,2\pi]} \left| \frac{e^{\frac{1}{\Delta} \cos(y-\mu)}}{2 \pi I_0(\frac{1}{\Delta})} 
- \frac{1}{\sqrt{2\pi \Delta}} \exp \left( -\frac{(y-\mu)^2}{2\Delta}\right)
 \right| =  0.$$

Then, one way to think about this is given $\underline{Y}_{[0:t]}=\underline{y}_{[0:t]}$, the random variable $Y_{i,t+\Delta}$ has a von Mises distribution with $\mu = y_{i,t} + \Delta \parenth{\omega_i + \sum_{j=1}^N \kappa_{i,j} \sin(y_{j,t}-y_{i,t}-\mu_{ij})}$ and inverse variance $\kappa$ proportional to $\frac{1}{\sigma^2 \Delta}$.
An autoregressive model of this form has been used (albeit without the $\kappa_{i,j} \sin(y_{j,t}-y_{i,t}-\mu_{ij})$ term) for phase time series estimation \cite{matsuda2017time}.

\vspace{1em}

\section{Stochastic Modeling of the Kuramoto Dynamics}
We now will consider a discrete time stochastic model that captures properties of the SDE for Kuramoto von Mises. Suppose we have a matrix of $N$ vectors over time, each of which are length $T$: 
\[
\underline{Y}_{1:T} = [y_{1,1}, \ldots, y_{1,T}; y_{2,1}, \ldots, y_{2,T}; \ldots y_{N,1}, \ldots, y_{N,T}] \in \mathbb{R}^{N \times T}
\]

From here on out, we will use $t$ as an integer index where we have sampled the continuous time process every $\Delta$ seconds. 
We now arrive at a ``spatial conditional independence'' property that builds directly upon Remark~\ref{remark:SDE:conditional-indepdnenece}:
\begin{remark} \label{remark:spatial-conditional-independence}
$(Y_{1,t},\ldots,Y_{N,t})$ are statistically independent given $\underline{Y}_{1:t-1}$.
\end{remark}
Then note that from the chain rule of conditional probability:
\begin{eqnarray}
f_{\underline{Y}_{1:T}} (\underline{y}_{1:T}) &=& f_{\underline{Y}_{1}}(\underline{y}_{1})  \prod_{t=1}^{T-1} f_{\underline{Y}_{t+1} | \underline{Y}_{1:t}} \parenth{\underline{y}_{t+1} | \underline{y}_{1:t}} \nonumber \\
															  &=&  \parenth{\prod_{i=1}^N f_{Y_{i,1}}(y_{i,1})}  \prod_{t=1}^{T-1} \parenth{\prod_{i=1}^N f_{Y_{i,t+1} | \underline{Y}_{1:t}} \parenth{y_{i,t+1} | \underline{y}_{1:t}}} \label{eqn:jointdistribution:b}
\end{eqnarray}
where \eqref{eqn:jointdistribution:b} follows from Remark~\ref{remark:spatial-conditional-independence}.

Further, since we discretize time into bins of length $\Delta$ to arrive at this discrete time model, we have: 
\begin{align}
f_{Y_{i,t+1} | \underline{Y}_{1:t}} \left(\cdot \mid \underline{y}_{1:t}\right)
  &= \text{vonMises}\!\left(\mu= y_{i,t} + \Delta \left(\omega_i +  \sum_{j=1}^N \kappa_{ij} \sin(y_{j,t}-y_{i,t}-\mu_{ij})\right); \nu = \tfrac{1}{\Delta \sigma^2} \right) \nonumber \\
  &= \text{vonMises}\!\left(\mu=  y_{i,t} +\Delta \left(\omega_i + \sum_{j=1}^N
      \kappa^c_{ij} \sin\!\left( y_j(t)-y_i(t)\right)
      + \kappa^s_{ij} \cos\!\left( y_j(t)-y_i(t)\right) \right); \nu = \tfrac{1}{\Delta \sigma^2} \right) \label{eqn:kur_vonmises_modified}
\end{align}
where \eqref{eqn:kur_vonmises_modified} comes from \eqref{eqn:kur_td}.

\newcommand{\VMKMa}{y_{i,t+1}- y_{i,t} - \omega_i \Delta}
\newcommand{\VMKMb}{\Delta \sum_{j=1}^N \kappa_{i,j} \sin(y_{j,t}-y_{i,t})}
\newcommand{\VMKMbb}{ \sum_{j=1}^N \kappa_{i,j} \sin(y_{j,t}-y_{i,t})}

\newcommand{\VMKMlag}{y_{i,t+1}- y_{i,t} - \omega_j \tau_{ij}}



We define the following terms for brevity:
\newcommand{\Ait}{A_{i,t}}
\newcommand{\Bit}{B_{i,t}}
\newcommand{\uit}{u_{i,t}}

\begin{align}\label{eqn:terms:brevity}
\Ait &\triangleq \sin \parenth{\VMKMa}\\
\Bit &\triangleq \cos \parenth{\VMKMa}\\
\underline{S}_{i,t} &\triangleq [\sin(y_{1,t}-y_{i,t}), \ldots, \sin(y_{N,t}-y_{i,t})]^T \in \mathbb{R}^N\\
\underline{C}_{i,t} &\triangleq [\cos(y_{1,t}-y_{i,t}), \ldots, \cos(y_{N,t}-y_{i,t})]^T \in \mathbb{R}^N\\
\underline{\kappa}_{i}^s &\triangleq [\kappa_{i1}^{s}, \ldots, \kappa_{iN}^{s}]^\top \in \mathbb{R}^N \\
\underline{\kappa}_{i}^c &\triangleq [\kappa_{i1}^{c}, \ldots, \kappa_{iN}^{c}]^\top \in \mathbb{R}^N \\
u_{i,t} &\triangleq \Delta \left(\sum_{j=1}^{N} \kappa_{ij}^c \sin(y_j(t) - y_i(t)) + \kappa_{ij}^s \cos(y_j(t) - y_i(t))\right) \\ &= \Delta(\underline{\kappa}_i^{c\top}\underline{S}_{i,t}+\underline{\kappa_{i}}^{s\top}\underline{C}_{i,t})
\end{align}

With this, we have:
\begin{eqnarray} \log P_{Y_{i,t+1} | \underline{Y}_{1:t}} \parenth{y_{i,t+1} | \underline{y}_{1:t}; \kappa} &=& \nu\cos\parenth{y_{i,t+1}-y_{i,t}-\omega_i\Delta - u_{i,t}} - \log (2 \pi I_0(\nu))\nonumber \\ &=& \nu \sin \parenth{\VMKMa} \sin \parenth{u_{i,t}} \nonumber \\ &+& \nu \cos \parenth{\VMKMa} \cos\parenth{u_{i,t}} \nonumber \\ &-& \log (2 \pi I_0(\nu)) \label{eqn:ll_delay_trig_id}\\
&=&\nu A_{i,t}\sin(u_{i,t}) + \nu B_{i,t}\cos(u_{i,t})-\log (2 \pi I_0(\nu))
\label{eqn:loglikelihooddelay} \end{eqnarray}
where in \eqref{eqn:ll_delay_trig_id}, we used the following trigonometric identity:
\beqa
\cos(a-b)=\sin a \sin b + \cos a \cos b   \label{eqn:trigidenttity}
\eeqa

\newcommand{\uitfull}{\Delta(\underline{\kappa}_i^{c\top}\underline{S}_{i,t}+\underline{\kappa_{i}}^{s\top}\underline{C}_{i,t})}

\newcommand{\uitnodel}{\underline{\kappa}_i^{c\top}\underline{S}_{i,t}+\underline{\kappa_{i}}^{s\top}\underline{C}_{i,t}}

\begin{remark}\label{remark:trigApproximation}
Now note that if we sample at high data rates, in which $\Delta \ll 1$, then for any bounded $\kappa$ matrix, it is clear that for sufficiently small $\Delta$, we have that 
\begin{equation}
|u_{i,t}| \ll 1
\end{equation}

Then we can use Taylor approximation for $\sin(u)$ and $\cos(u)$ when $u \simeq 0$ (e.g. $|u| \ll 1$):
\beqa
\sin u \simeq u \\
\cos u \simeq 1-\half u^2 .
\eeqa
The above follows from the approximation $f(u) \simeq f(0) + f'(0)u + \frac{f''(0)}{2}u^2$ for $u \simeq 0$ and from the fact that $\sin(0)=0, \sin'(0)=1, \sin''(0)=0,\cos(0)=1,\cos'(0)=0, \cos''(0)=-1$.
\end{remark}

As such, with this approximation that applies when discretizing continuous time in analogy with point processes \cite{smith2003estimating}, we have:

\begin{eqnarray}
-\log  f_{Y_{i,t+1} | \underline{Y}_{1:t}} \parenth{y_{i,t+1} | \underline{y}_{1:t}; \kappa} &=&  -\nu \Ait \sin \uit - \nu \Bit \cos \uit + \log (2 \pi I_0(\nu)) \nonumber \\
	&\simeq& -\nu \Ait \uit - \nu \Bit \parenth{1-\half \uit^2} + \log (2 \pi I_0(\nu)) \nonumber \\
	&=& -\nu \Ait \parenth{\uitfull} - \nu \Bit \parenth{1-\half \parenth{\uitfull}^2} \nonumber \\
	&+& \log (2 \pi I_0(\nu)) \label{eqn:loglikelihood:approximation:continoustime}
\end{eqnarray}
Note that this approximation becomes more exact as $\Delta$ approaches $0$.  In practical terms, this suggests that if we have data sampled at a rate much faster than the underlying oscillation $\omega$, then it is likely a reasonable approximation. Analogous approximations, involving the transition from continuous time to discrete time, have been utilized in point process modeling \cite{smith2003estimating}.
We can write the formulation for maximum likelihood estimation of $\kappa$ as the following:
\begin{eqnarray}
\hat{\kappa}_{\text{ML}} &=& \arg \min_\kappa -\log f_{\underline{Y}_{1:T}} (\underline{y}_{1:T}; \kappa) \nonumber \\
						&=& \arg \min_\kappa -\log \brackets{  \prod_{t=1}^{T-1} \parenth{\prod_{i=1}^N f_{Y_{i,t+1} | \underline{Y}_{1:t}} \parenth{y_{i,t+1} | \underline{y}_{1:t}; \kappa}}} \nonumber\\
						&=& \arg \min_\kappa \sum_{t=1}^{T-1} \sum_{i=1}^N -\log f_{Y_{i,t+1} | \underline{Y}_{1:t}} \parenth{y_{i,t+1} | \underline{y}_{1:t}; \kappa} \nonumber\\
						&=& \arg \min_\kappa \sum_{t=1}^{T-1} \sum_{i=1}^N -\nu \Ait \parenth{\uitfull} +\half \nu \Bit \parenth{\uitfull}^2 \label{eqn:MLestimation:aaa}\\
						&=& \arg \min_\kappa \sum_{t=1}^{T-1} \sum_{i=1}^N -\frac{1}{\Delta \sigma^2} \Ait \parenth{\uitfull} +\half \frac{1}{\Delta \sigma^2} \Bit \parenth{\uitfull}^2 \label{eqn:MLestimation:aa}\\
						&=& \arg \min_\kappa \sum_{t=1}^{T-1} \sum_{i=1}^N - \Ait \parenth{\uitnodel} +\half \Delta  \Bit \parenth{\uitnodel}^2 .\label{eqn:MLestimation:a}
\end{eqnarray}

We further simplify this maximum likelihood estimation in closed form.  Define $J(\kappa)$ as:
\begin{eqnarray}\label{eqn:obj_func}
J(\kappa) \triangleq  \sum_{i=1}^N \sum_{t=1}^{T-1} -\Ait \parenth{\uitnodel} +\half \Delta \Bit \parenth{\uitnodel}^2  
\end{eqnarray}

\newcommand{\vkappa}{\underline{\kappa}}
\newcommand{\vS}{\underline{S}}
Note that if we define $\vkappa_i$ as $[\kappa_{i,1},\ldots,\kappa_{i,N}]^T$, then from the outer sum over $i$ it is clear that this problem is separable in $\vkappa_i$:
\begin{eqnarray}
J(\kappa) &=&  \sum_{i=1}^N J_i(\vkappa_i) \\
J_i(\vkappa_i) &\triangleq&  -\Ait \parenth{\uitnodel} +\half \Delta \Bit \parenth{\uitnodel}^2  
\end{eqnarray}
 
Define 
\begin{align}
\begin{aligned}
\underline{SC}_{i,t} 
&\triangleq
\begin{bmatrix}
\underline{S}_{i,t} \\
\underline{C}_{i,t}
\end{bmatrix}
\end{aligned}
\hspace{2em}
\begin{aligned}
\underline{\kappa}_i^{cs} 
&\triangleq
\begin{bmatrix}
\underline{\kappa}_{i}^{c} \\
\underline{\kappa}_{i}^{s}
\end{bmatrix}
\end{aligned}
\end{align}

Then we have the following expression as the objective function.
\begin{eqnarray}
J_i(\underline{\kappa}_i) = \sum_{t=1}^{T-1} -A_{i,t} \underline{\kappa}_i^{cs\top} \underline{SC}_{i,t}+\frac{\Delta}{2}B_{i,t}\underline{\kappa}_i^{cs\top} \underbrace{\underline{SC}_{i,t} \underline{SC}_{i,t}^\top}_{Z}\underline{\kappa}_i^{cs}
\end{eqnarray}
Notice this is a quadratic form with respect to $\underline{\kappa}_i^{cs}$. The matrix $Z$ must be positive semidefinite because it is an outerproduct. Note that the $i$th entry of $\vS_{i,t}$  is $0$ and so finding $\kappa_{i,i}$ as it stands now is not possible - it is not observable.  We add a quadratic cost on the $i^{th}$ term in $\underline{S}_{i,t} \text{ and } \underline{C}_{i,t}$, encouraging it be $0$. After defining 
$E_i = \begin{bmatrix}
\underline{e}_i \\
\underline{e}_i
\end{bmatrix}$, we have our objective function as:

\renewcommand{\tJ}{\tilde{J}}
\newcommand{\ve}{\underline{e}}
\newcommand{\vb}{\underline{b}}
\newcommand{\vzero}{\underline{0}}

\beqa
\tJ_i(\vkappa_i) &=& J_i(\vkappa_i)  + \half \Delta (\vkappa_{i,i}^{c\,2} + \vkappa_{i,i}^{s \, 2}) \label{eqn:tJi:addedterm}\\
			   &=& 	\sum_{t=1}^{T-1} -A_{i,t} \underline{\kappa}_i^{cs\top} \underline{SC}_{i,t}+\frac{\Delta}{2}B_{i,t}\underline{\kappa}_i^{cs\top} \underline{SC}_{i,t} SC_{i,t}^\top\underline{\kappa}_i^{cs} + \vkappa_i^{cs \top} \underline{E}_i \underline{E}_i^{\top} \vkappa_i^{cs} \nonumber
\eeqa

and so taking the gradient and setting to $0$, we have
\beqa
\vzero &=& \nabla \tJ_i(\vkappa_i) \\
	   &=& \left(\sum_{t=1}^{T-1} - \Ait  \underline{SC}_{i,t} + \Delta \Bit \underline{SC}_{i,t} \underline{SC}_{i,t}^T \underline{k}_i^{cs} \right) + \Delta \underline{E}_i \underline{E}_i^T \underline{k}_i^{cs} \\
\eeqa
If we let
\begin{subequations}
\label{eqn:defn:bandM}
\beqa
\vb_i &\triangleq& \sum_{t=1}^{T-1}  \Ait  \underline{SC}_{i,t} \label{eqn:defn:bi}\\
M_i &=& \Delta \parenth{\sum_{t=1}^{T-1}  \Bit \underline{SC}_{i,t} \underline{SC}_{i,t}^T + \underline{E}_i \underline{E}_i^T} \label{eqn:defn:Mi}
\eeqa
\end{subequations}
we have our final form:
\beqa
\hat{\underline{\kappa}}_i^{cs} &=& M^{-1}_i \vb_i \label{eqn:kappahat:matrixinverse}
\eeqa

\begin{remark} \label{remark:linalg_est}
Estimating $\hat{\underline{\kappa}}_i^{cs}$ reduces to linear algebraic operations that can be solved efficiently without requiring a convex optimization solver. In practice, we use a standard Lasso solver to induce sparsity in $\hat{\underline{\kappa}}_i^{cs}$. With the estimated $\hat{\underline{\kappa}_i^{cs}}$, we can compute $\kappa_{ij}$ and $\mu_{ij}$ using standard trigonometric operations:
\begin{eqnarray}
\kappa_{ij} &=& \sqrt{\kappa_{ij}^{s2} + \kappa_{ij}^{c2}} \\ 
\mu_{ij} &=& \arctan \parenth{\frac{-\kappa_{ij}^s}{\kappa_{ij}^c}}
\end{eqnarray}
\end{remark}

\begin{remark} 
In certain scenarios, one can be in a regime with instantaneous coupled interaction without time delays. In those cases, we can easily follow the same approach starting with the original Kuramoto formulation in \eqref{eqn:original:Kuramoto}.
\end{remark}

\subsection{Estimation in Symmetric Connectivity}
Although our proposed approach provides general framework without requiring equilibrium assumptions, we can still enforce equilibrium with symmetric connectivity structure within the estimation procedure. Assuming symmetric connectivity, we know $\kappa_{ij} = \kappa_{ji}$. Then, instead of estimating $N(N-1)$ number of $\kappa_{ij}$, we estimate $\frac{N(N-1)}{2}$ number of $\kappa_{ij}$. To start, we define the following:
\begin{subequations} \label{eqn:terms:symm}
\beqa
c_{ij}^s &\triangleq& \kappa_{ij}^s \text{ for $i<j$} \nonumber \\
c_{ij}^c &\triangleq& \kappa_{ij}^c \text{ for $i<j$} \nonumber \\
\psi_{i,j}^s(t) &\triangleq& \sin(y_{j,t}-y_{i,t}) \nonumber \\
\psi_{i,j}^c(t) &\triangleq& \cos(y_{j,t}-y_{i,t})\nonumber \\
x_{i,t} &\triangleq& \sum_{j=1}^{N} \kappa_{ij}^c \psi_{i,j}^s(t) +\kappa_{ij}^s\psi_{ij}^c(t) 
\label{eqn:def_symm_conn}
\eeqa
\end{subequations}

We can rewrite (\ref{eqn:obj_func}) as:
\begin{eqnarray}\label{eqn:substituted_x}
J(\kappa) = \sum_{i=1}^{N}\sum_{t=1}^{T-1}-A_{i,t}x_{i,t}+\frac{\Delta}{2}B_{i,t}x_{i,t}^2
\end{eqnarray}

We split the summation in \eqref{eqn:def_symm_conn} into two and write it with respect to $c_{ij}^s \text{ and } c_{ij}^s$ :
\begin{eqnarray}
x_{i,t} &=& \sum_{j>i}^{N} c_{ij}^c \psi_{ij}^s(t) +c_{ij}^s\psi_{ij}^c(t) + \sum_{j<i}^{N} c_{ji}^c \psi_{ij}^s(t) +c_{ji}^s\psi_{ij}^c(t) \\
&=& \sum_{j>i}^{N} c_{ij}^c \psi_{ij}^s(t) +c_{ij}^s\psi_{ij}^c(t) + \sum_{j<i}^{N} -c_{ji}^c \psi_{ji}^s(t) +c_{ji}^s\psi_{ji}^c(t) \label{eqn:flip_edge} \\
&=& \sum_{j>i}^{N} c_{ij}^c \psi_{ij}^s(t) + \sum_{j<i}^{N} -c_{ji}^c \psi_{ji}^s(t) + \sum_{j>i}^{N}c_{ij}^s\psi_{ij}^c(t) + \sum_{j<i}^{N}c_{ji}^s\psi_{ji}^c(t)
\end{eqnarray}

where $c_{ji}^c$, $\psi_{ji}^s$, $c_{ji}^s$, and $\psi_{ji}^c$ from \eqref{eqn:flip_edge} respect the order since $c_{ij}^c \text{ and } c_{ij}^s$ are only defined in pairs $(i,j)$ where $i<j$, and also $\psi_{ij}^s = -\psi_{ji}^s$ and $\psi_{ij}^c = \psi_{ji}^c$. Through this reparameterization, we effectively reduced the number of parameters by half. Now, consider the vectorized form of $x_{i,t}$ where each row entry represents a pair of node and time point $(i,t)$:
\begin{eqnarray}
\underline{x} &\triangleq& [x_{1,1} \,  ,  \, x_{2,1} \,, \, ... \, , \, x_{N,1} \, , \, x_{1,2} \, , ... ,\, x_{2,2}\, , \, ... \, ,\, x_{N,T-1}]^{\top} \in \mathbb{R}^{N (T-1)}
\end{eqnarray}
We also consider the vectorized form of $c_{i,j}^c \text{ and } c_{i,j}^s$ where each row entry is an edge between two oscillators $i$ and $j$ where $i<j$ :
\begin{eqnarray}
\underline{c}^c \triangleq [c_{1,2}^c \,,c_{1,3}^c \, , \, ... \, , c_{1,N}^c \,, c_{2,3}^c \,,\, ... \,, c_{N-1,N}^c]^{\top} \in \mathbb{R}^{\frac{N(N-1)}{2}} \\
\underline{c}^s \triangleq [c_{1,2}^s \,,c_{1,3}^s \, , \, ... \, , c_{1,N}^s \,, c_{2,3}^s \,,\, ... \,, c_{N-1,N}^s]^{\top} \in \mathbb{R}^{\frac{N(N-1)}{2}}
\end{eqnarray}

Then, we can write $\underline{x}=F\underline{c}^c + G\underline{c}^s$ , where the rows of $F$ and $G$ correspond to $(i,t)$, and columns correspond to an edge $(p,q)$ where $p<q$. Then each entry of $F_{(i,t),(p,q)}$ and $G_{(i,t),(p,q)}$  is the following,
\begin{eqnarray}
F_{(i,t),(p,q)} &=&
\begin{cases}
+\psi_{pq}^s(t) & \text{if } i = p, \\
-\psi_{pq}^s(t) & \text{if } i = q, \\
0 & \text{otherwise}.
\end{cases}, \text{ where } F \in \mathbb{R}^{N(T-1) \times \frac{N(N-1)}{2}} \\
G_{(i,t),(p,q)} &=&
\begin{cases}
+\psi_{pq}^c(t) & \text{if } i = p \text{ or }q, \\
0 & \text{otherwise}.
\end{cases}, \text{ where } G \in \mathbb{R}^{N(T-1) \times \frac{N(N-1)}{2}}
\end{eqnarray}

After defining:
\begin{subequations} \label{eqn:terms:symm}
\beqa
H &\triangleq& [F \; G] \in \mathbb{R}^{N(T-1) \times N(N-1)} \nonumber \\
\underline{c} &\triangleq& \begin{bmatrix}
\underline{c}^{c} \\
\underline{c}^{s} 
\end{bmatrix} \in \mathbb{R}^{N(N-1)} \nonumber\\
\underline{a} &\triangleq& [A_{1,1},\, A_{1,2},\, ... , A_{1,T-1},\, A_{2,1},\, A_{2,2},\, ..., 
, A_{N,T-1}]^{\top} \in \mathbb{R}^{N(T-1)} \nonumber\\
W &\triangleq& \text{diag}(\Delta B_{i,t}) \in \mathbb{R}^{N(T-1) \times N(T-1)} \nonumber
\eeqa
\end{subequations}

we have that $\underline{x} = H\underline{c}$, then we rewrite objective function in \eqref{eqn:substituted_x} as:
\begin{eqnarray}
J(\underline{c}) = -\underline{a}^{\top}H\underline{c}+\frac{1}{2}\underline{c}^{\top}H^{\top}WH\underline{c}
\end{eqnarray}

Although it is possible to enforce symmetric structure in our method as shown here, we demonstrate that even without this reparameterization, our approach yields estimates comparable to methods assuming equilibrium in section 5.
\vspace{1em}

\section{Estimation of $\sigma^2$}
Our approach models the coupled oscillator system probabilistically, enabling explicit characterization of uncertainty. Note that $\nu=\frac{1}{\sigma^2 \Delta}$.  As $\Delta \to 0$, $\nu \to \infty$ and the Bessel function becomes:
\begin{eqnarray*}
I_0(\nu) &=& \frac{e^{\nu}}{\sqrt{2\pi\nu}}\left(1+\frac{1}{8\nu}\right) \simeq \frac{e^{\nu}}{\sqrt{2\pi\nu}} \\
 -\log I_0(\nu)  &=&-\nu + \frac{1}{2} \log(2 \pi \nu) 
\end{eqnarray*}
As such, when we write down $NLL(\kappa,\nu)$, including the Bessel function term and eliminating constant terms, we have:
\begin{eqnarray}
NLL(\kappa,\nu) &=&  \sum_{t=1}^{T-1} \sum_{i=1}^{N} \left( -\nu \Ait u_{i,t} -  \nu \Bit (1- \half u_{i,t}^2) + \log (I_0(\nu)) \right) \\
	&=&  \sum_{t=1}^{T-1}\sum_{i=1}^{N} \left( -\nu \Ait u_{i,t} -\nu\Bit+ \half \nu \Bit u_{i,t}^2 - \nu + \frac{1}{2} \log(2 \pi \nu) \right)
\end{eqnarray}
This means that 
\begin{eqnarray}
0 &=& \frac{\partial}{\partial \nu }NLL(\hat{\kappa},\nu) \\
  &=& \sum_{t=1}^{T-1} \sum_{i=1}^{N} \left(-\Ait u_{i,t} - \Bit  + \half \Bit u_{i,t}^2 -1 + \frac{1}{2\nu} \right) \\
  &=& \sum_{t=1}^{T-1} \sum_{i=1}^{N} \left( -\Ait u_{i,t} -  \Bit + \half \Bit u_{i,t}^2 -1 + \half \Delta \sigma^2 \right) \\
  &=& \sum_{t=1}^{T-1}  \sum_{i=1}^{N} \left(  -\Ait \parenth{\underline{\kappa}_i^{c\top}\underline{S}_{i,t}+\underline{\kappa_{i}}^{s\top}\underline{C}_{i,t}}- \frac{1}{\Delta}\Bit + \half \Delta \Bit \parenth{\underline{\kappa}_i^{c\top}\underline{S}_{i,t}+\underline{\kappa_{i}}^{s\top}\underline{C}_{i,t}}^2 -\frac{1}{\Delta} + \half \sigma^2 \right ) \\
  &=& \sum_{t=1}^{T-1}  \sum_{i=1}^{N} -\Ait \parenth{\underline{\kappa}_i^{c\top}\underline{S}_{i,t}+\underline{\kappa_{i}}^{s\top}\underline{C}_{i,t}} + \half \Delta \Bit \parenth{\underline{\kappa}_i^{c\top}\underline{S}_{i,t}+\underline{\kappa_{i}}^{s\top}\underline{C}_{i,t}}^2 \\
  &+& \sum_{t=1}^{T-1}  \sum_{i=1}^{N} \left( -\frac{1}{\Delta}\Bit-\frac{1}{\Delta}+\frac{\sigma^2}{2} \right) \\
  &=& J(\kappa) -\frac{1}{\Delta} \sum_{t=1}^{T-1}  \sum_{i=1}^{N} \Bit - \frac{(T-1)N}{\Delta} + \frac{(T-1)N}{2}\sigma^2
\end{eqnarray}
Now we have the final form as:
$$\hat{\sigma}^2 =  \frac{2}{(T-1)N} \left(   \sum_{t=1}^{T-1}  \sum_{i=1}^{N} \frac{\Bit}{\Delta}  - J(\kappa) \right) + \frac{2}{\Delta} $$

\section{Experiments}
In this section, we evaluate the robustness of our proposed approach using both synthetic data generated from a generative model and electrophysiological recordings.

\subsection{Graph Recovery in Symmetric, Equilibrium State}
\begin{figure}[h]  
    \centering
    \includegraphics[width=0.8\textwidth]{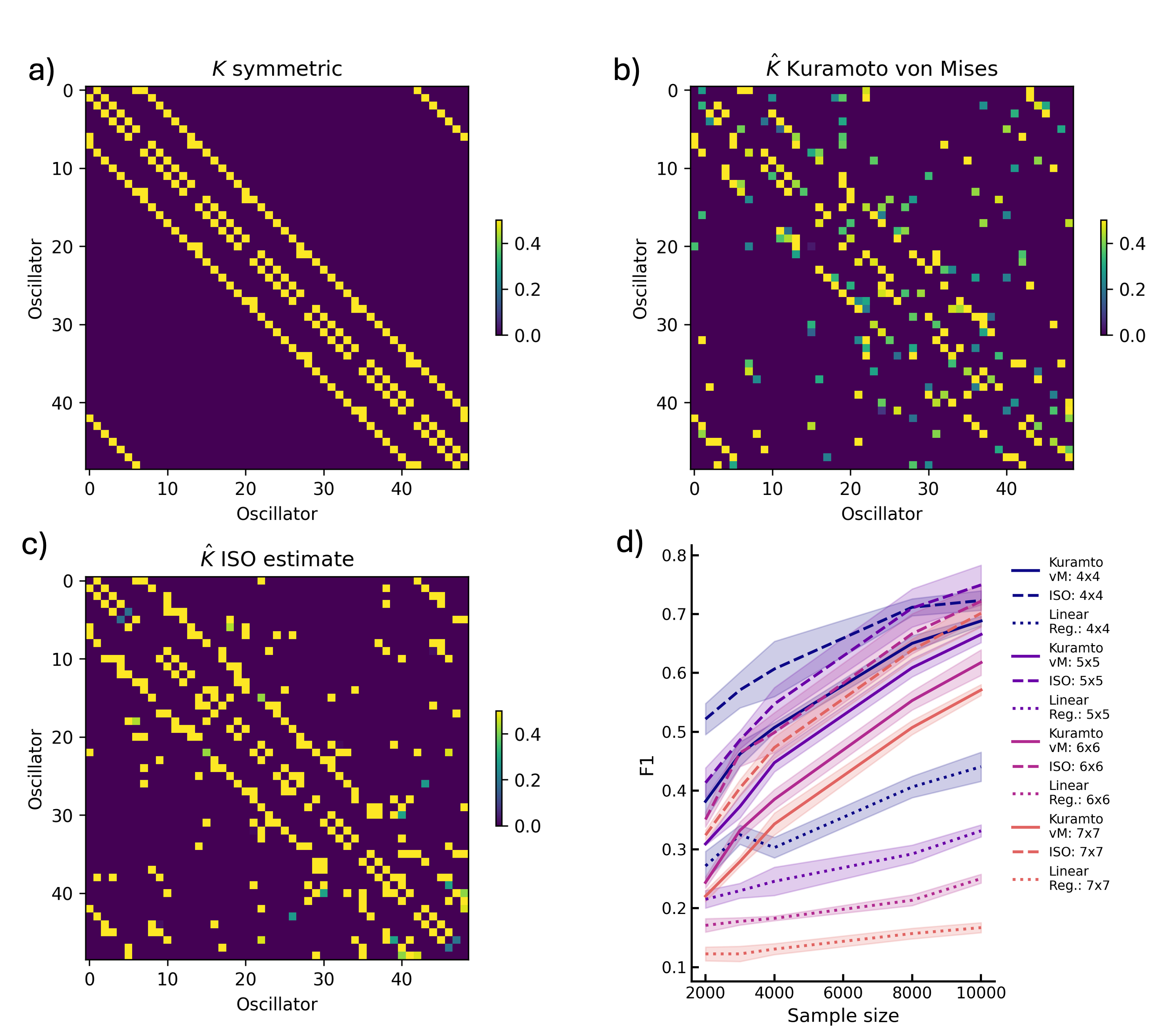}
    \caption{Model performance in graph recovery of a system of coupled oscillators under symmetric connectivity. (a) True symmetric $K$ of 49 oscillators organized in a $7\times7$ array that was used to generate data. (b) estimated $K$ using our proposed MLE algorithm. (c) estimated $K$ using the ISO algorithm that assumes an equilibrium Boltzmann distribution \cite{perley_coleman_2025_phase}. (d) F1-score with respect to the true graph structure in varying dimensions. Solid line shows the performance of the proposed Kuramoto von Mises algorithm. Dashed line shows the performance of the ISO method from from \cite{perley_coleman_2025_phase}. Dotted line represents the performance of linear regression. Our model provides estimates comparable to those of the ISO approach.}
    \label{fig:figure1}
\end{figure}

We test our model on simulated data with varying dimensionality to assess its ability to recover high-dimensional graph structure of a system of coupled oscillators with symmetric connectivity. Each oscillator $i$ on the grid has four neighbors: above, below, to the left, and to the right. In cases where oscillator is located on an edge, the edge wraps around to the other side, generating a symmetric connectivity (Figure~\ref{fig:figure1}a). Oscillators have background oscillation $\omega_i \sim \mathcal{N}(\pi/2, 0.1)$. Sampling interval $dt$ was set to 10 ms, and each step of the dynamic contained gaussian noise with noise constant $\sigma = 1$ from \eqref{eqn:kuramoto_noise}. Phase delay constant $\tau_{ij}$ was set 100 ms.

We implement a two-step model fitting procedure for our model and the ISO approach from \cite{perley_coleman_2025_phase} as a comparison. We also implement a simple baseline using Lasso regression to estimate $\underline{\kappa}_i^{s}$ and $\underline{\kappa}_i^{c}$ from \eqref{eqn:kur_td}. First, we fit the model with an $\ell_{1}$ regularizer to obtain a sparse estimate of the connectivity. Second, we select the top $x\%$ of the $\kappa_{ij}$'s and refit an unregularized model using only the oscillators identified as connected. We choose the regularization strength and the threshold $x$ via grid search.  The regularization strength when using the ISO approach was selected using theoretical results from the original work \cite{perley_coleman_2025_phase}. 

Both our approach and the ISO method recover the underlying symmetric connectivity matrix $K$ (Figure~\ref{fig:figure1}b-c). We evaluate graph retrieval performance using the F1-score across varying training set sizes (Figure \ref{fig:figure1}d). ISO performs slightly better than our approach. This result is expected, since the ISO method assumes symmetric connectivity, which happens to align with the structure of the generated data (Table \ref{tab:estimation_comparison}). Nonetheless, we highlight that even though we do not assume symmetric and Boltzmann equilibrium assumptions, our approach provides an accurate estimate of the symmetric connectivity. A comparison with standard linear regression shows that our approach performs better across all dimensions and training data sizes.

\begin{table}[h]
\centering
\renewcommand{\arraystretch}{1.3}
\begin{tabular}{p{3.5cm} c c c}
\toprule
Modeling Assumptions \\ \& Capabilities & Kuramoto–von Mises & ISO & Linear regression \\
\hline
Generalizable to \\ asymmetric interactions? & O & X & O \\
Generalizable to \\ nonequilibrium settings? & O & X & O \\
Works in wide-band $\underline{\omega}$? & X & O & X \\
Scales to high-dimensional setting? & O & O & X \\
\bottomrule
\end{tabular}
\vspace{6pt} %
\caption{Properties of different estimation procedures}
\label{tab:estimation_comparison}
\end{table}

\subsection{Graph Recovery in Asymmetric, Non-equilibrium State}
\begin{figure}[h]  
    \centering
    \includegraphics[width=0.8\textwidth]{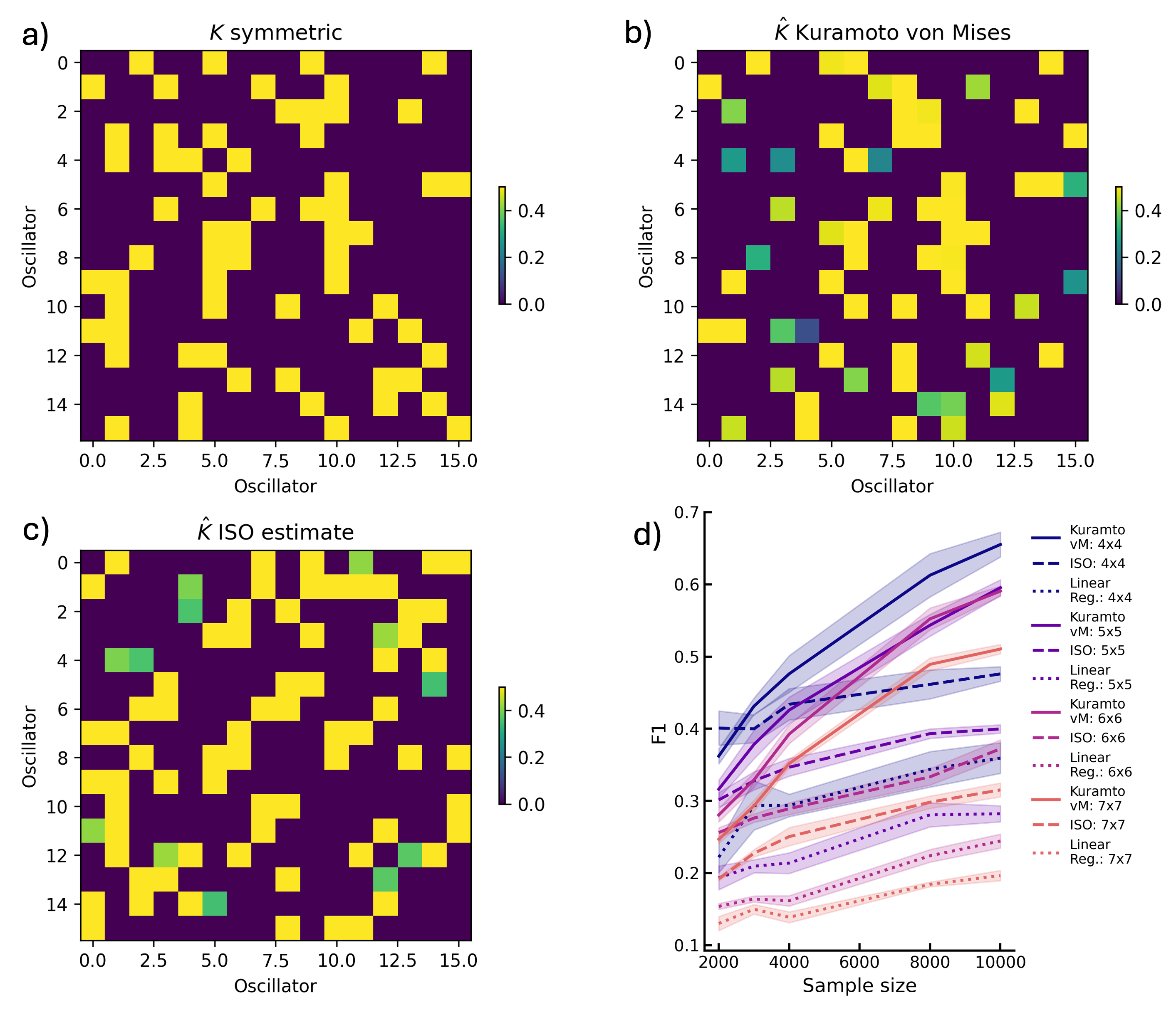}
    \caption{Model performance in graph recovery of a system of coupled oscillators with asymmetric interactions. (a) True asymmetric $K$ of 16 oscillators organized in a $4\times4$ array that was used to generate data. (b) estimated $K$ using our proposed MLE algorithm. (c) estimated $K$ using the ISO algorithm that assumes an equilibrium Boltzmann distribution \cite{perley_coleman_2025_phase}. (d) F1-score with respect to the true graph structure in varying dimensions. Solid line shows the performance of the proposed MLE algorithm. Dotted line represents the performance of the ISO algorithm. Our proposed MLE approach performs better than ISO in all array sizes.}
    \label{fig:figure2}
\end{figure}

Although ISO has better overall performance in symmetric systems, many physical systems exhibit asymmetric interactions as previously mentioned. To generate kuramoto dynamics with asymmetric interactions, we shuffled $K$  and generated asymmetric connectivities (i.e., $\kappa_{ij} \neq \kappa_{ji}$). The shuffled $K$  has the same degree of neighborhood and connectivity strength as before in the symmetric $K$. Estimated $K$ from our MLE approach captures the asymmetric structure while the ISO method does not (Figure~\ref{fig:figure2}b-c). Our approach learns the underlying connectivity as the data size increases, as evidenced by the increasing F1-score, whereas the F1-score from ISO plateaus starting at lower training data size. (Figure~\ref{fig:figure2}d, Table \ref{tab:estimation_comparison}). Similar to the symmetric case, our approach performs better across all dimensions and sample sizes compared to standard linear regression. 

\subsection{Likelihood Ratio Testing to Classify Traveling Waves and Gastric Waves}
\begin{figure}[h]  
    \centering
    \includegraphics[width=0.8\textwidth]{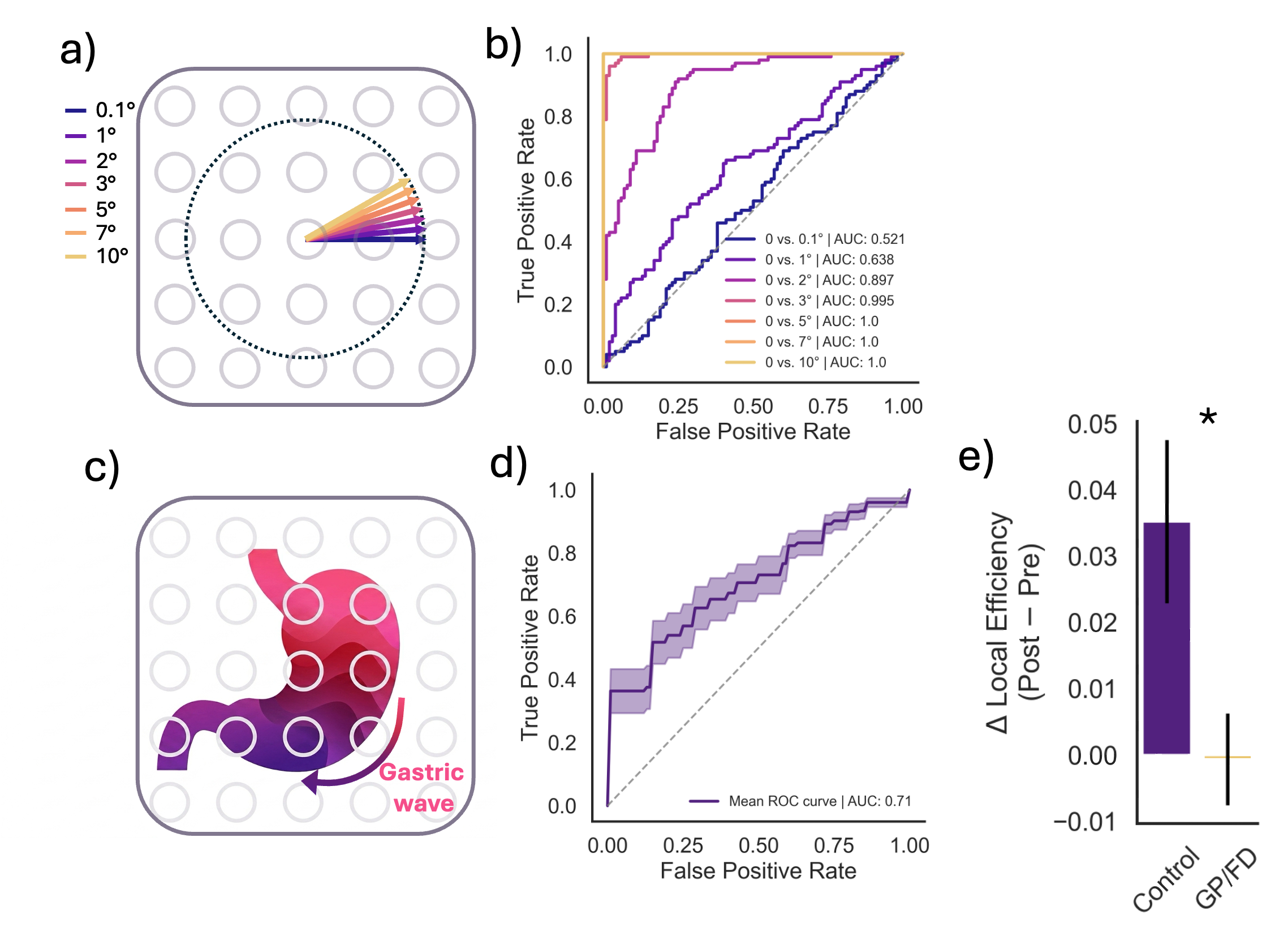}
    \caption{Traveling wave classification. (a) Traveling plane waves propagating in varying directions were generated. Each circle is an oscillator, and each arrow with a different color represents a different wave propagation direction. (b) Receiver operating characteristic curve from a log-likelihood ratio test classifying traveling waves moving in different directions against a wave traveling at $0^\circ$. (c) Gastric wave that propagates from the top of the stomach towards the small intestine. These gastric waves become more prominent during digestive periods. (d) ROC in pre- vs. post-meal classification using our proposed approach. (e) Changes in local connectivity pre vs. post-meal in healthy controls and patients with gastric dysfunctions. In healthy controls, there is an increase in local connectivity during digestive periods compared to pre-meal conditions, consistent with the expected coordinated postprandial contractile patterns of gastric waves. However, in patients with gastroparesis or functional dyspepsia, the connectivity estimated by our model captures a pathological lack of change in connectivity between the pre- and post-meal states, potentially leading to abnormal digestion (p<0.05, t-test). } 
    \label{fig:figure3}
\end{figure}

In the context of neurophysiology, and many other physical systems, neurons (or oscillators) are organized in a way that often exhibit sequential activity and produce traveling waves. In this section, we simulate 2D planar traveling waves propagating in different directions using the following wave equation:
\[
u(x,y,t) = A\sin \left(k(x\cos(\theta)+y\sin(\theta)-\omega t + \phi_0) \right)
\]
where $A$ is the wave amplitude, $k$ is the wave number, $\theta$ is the propagation angle, $\omega$ is the angular frequency, and $\phi_0$ is the initial phase. In our simulation, we let $A=1$, $k=\pi/4$, $\omega = \pi$, with $x$ and $y$ each defined on a uniformly spaced $5\times 5$ grid over $[-3,3]$. Data was sampled every $dt=10\text{ ms}$. We simulate planar waves propagating in different directions by setting $\theta \in [0.1^{\circ}, \,1^{\circ}, \,2^{\circ}, \,3^{\circ}, \,5^{\circ}, \,7^{\circ}, \,10^{\circ}]$ and conduct log-likelihood (LL) ratio testing against a traveling wave propagating at $\theta=0^{\circ}$ (Figure~\ref{fig:figure3}a):
\[
LL(\underline{y}\,;\,\kappa_{0^{\circ}},\mu_{0^{\circ}})-LL(\underline{y}\,;\,\kappa_{\theta},\mu_{\theta}) \lessgtr \alpha
\]
where the LL was derived from \eqref{eqn:obj_func}. Each model was trained on 1000 samples and tested on 200 segments (100 segments on each propagation direction) consisting of 25 samples each, and the Receiver-Operating Curve (ROC) is shown (Figure~\ref{fig:figure3}b). As expected, the area under the curve (AUC) increases with greater differences in the direction of wave propagation, approaching 1 when comparing waves traveling to the right with those traveling  $3^{\circ}$ to the north. This result highlights the benefit of our model's probabilistic description of the phase distribution for hypothesis testing.

Next, we evaluate the applicability of our approach using high spatial resolution electrogastrography (EGG) data with 25 electrodes that was previously collected from healthy participants and patients with gastric dysfunctions at 5 Hz sampling rate \cite{gharibans2019spatial}. EGG is a noninvasive data modality capable of capturing electrical activity emanating from the neuromuscular cells of the stomach. As mentioned previously, these cells are known to organize their behavior to produce traveling waves, which are known to be more pronounced during digestive periods (Figure~\ref{fig:figure3}C). We test whether our proposed method can capture changes in gastric traveling waves before and after a meal using the same log-likelihood ratio test framework, after band-pass filtering the EGG signal to the relevant gastric frequency band (0.03–0.07 Hz) \cite{balasubramani2022gutbrain}. Since our model requires contiguous segments of data, we train on the first 75\% of the time series ( both pre- and post-meal periods) and partition the remaining 25\% into contiguous segments of 300 samples for testing. We add a 20 second buffer period, which is approximately one gastric cycle, in between train and test data to minimize information leakage from temporal adjacency. This setup evaluates the generalization of the trained model forward in time. 

Our approach provides a robust classification performance in classifying time segments before and after meals with AUC of 0.71 (Figure~\ref{fig:figure3}d). From the $\kappa$ connectivity matrix, we can observe connectivity patterns underlying gastric waves. We compute local efficiency using the connectivity matrix, a graph-theoretic measure that quantifies how efficiently information is exchanged within a node’s neighborhood after the node is removed \cite{latora2001efficient}. It is defined as follows:
\[
E_{\mathrm{loc}}(i) = \frac{1}{|N_i|(|N_i|-1)} \sum_{\substack{j,k \in N_i \\ j \neq k}} \frac{1}{d_{jk}^{(i)}}
\]
where $N_i$ represents the neighborhood of $i$, and $d_{jk}^{(i)}$ is the shortest path length between nodes $j$ and $k$ in the subgraph induced by $N_{i}$.  In control patients, there is an increase in local connectivity during digestive periods compared to pre-meal conditions, consistent with the expected coordinated contractile patterns of gastric waves for digestion (Figure ~\ref{fig:figure3}e). However, in patients with gastroparesis (GP) or functional dyspepsia (FD), the connectivity estimated by our model captures a pathological lack of change in connectivity between the pre- and post-meal states, potentially leading to abnormal digestion (p<0.05, t-test).

\subsection{Inferrence on Network Connectivity Patterns in intracranial electroencephalogram Data  }
\begin{figure}[h]  
    \centering
    \includegraphics[width=0.8\textwidth]{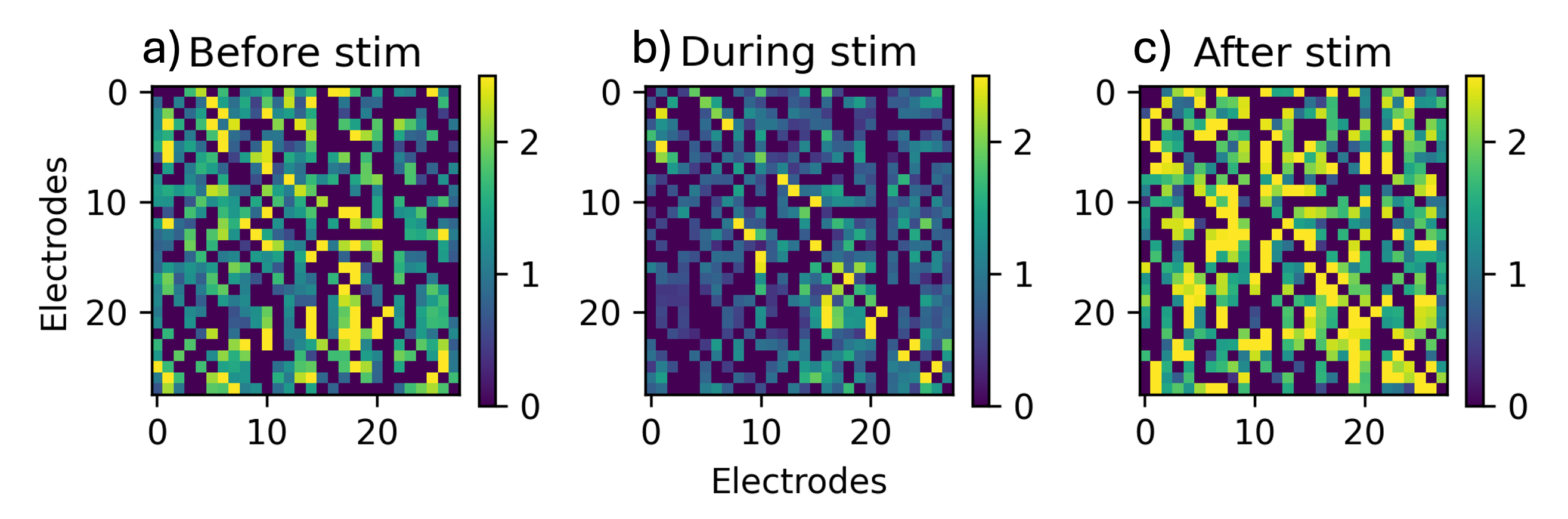}
    \caption{Network connectivity estimation in the orbitofrontal cortex in response to deep brain stimulation \cite{rao2018lof_stimulation_mood}. Our proposed approach captures changes in OFC connectivity across electrodes in response to DBS. a) Baseline connectivity matrix. b) Connectivity matrix during continuous 100 Hz DBS. c) Connectivity matrix after the stimulation. Network connectivity attenuates during stimulation relative to baseline and exhibits a rebound increase afterward.}
    \label{fig:figure4}
\end{figure}

In addition to gastric recordings, we test if our approach can be applied to intracortical recordings. We use a publicly available dataset of intracortical recordings from the orbitofrontal cortex of a patient undergoing continuous deep brain stimulation (DBS) at 1 mA and 100 Hz \cite{rao2018lof_stimulation_mood}. We focus on the theta frequency 4 – 8 Hz, as this frequency band had the biggest response to DBS. Details of their method of reconstructing theta frequency from the stimulation artifact can be found in their original work. Our approach is able to capture significant changes in $K$ connectivity before, during, and after the continuous DBS (Figure \ref{fig:figure4}a-c). $\kappa_{ij}$ entries decreased in response DBS compared to baseline (Figure \ref{fig:figure4}a,b), but showed an overall increase after stimulation (Figure \ref{fig:figure4}a,c). The original work showed decreased theta frequency band power during DBS compared to baseline, and increased after the stimulation \cite{rao2018lof_stimulation_mood}. The post-stimulation rebound in connectivity indicates a more strongly coupled state, in which each region receives convergent inputs from multiple sources, increasing its effective sensitivity to network activity. This enhanced input integration may, in turn, drive the observed increase in theta power after stimulation.

\vspace{1em}

\section{Discussion}

In this work, we propose a method to estimate the underlying dynamics of a system of coupled oscillators using the Langevin formulation, with phases modeled using a von Mises distribution. We show that our maximum likelihood estimation method has a closed form algebraic solution in the high sampling rate regime, a condition commonly satisfied by modern data acquisition systems, which makes it readily applicable in practice. Prior work has developed an algebraic formulation of the generative process underlying Kuramoto dynamics \cite{muller2021algebraic}. By lifting the phase into a complex-valued state space, the system admits a matrix representation amenable to linear algebraic operations, with the argument of the solution recovering the original Kuramoto dynamics. When combined with our approach for estimating the Kuramoto dynamics directly from noisy data, this yields a unified pipeline in which the coupling structure is first inferred and then analyzed within a linear algebraic framework, enabling scalable and interpretable analysis of high-dimensional coupled oscillator systems \cite{nguyen2023equilibria}.

Importantly, this framework does not rely on thermodynamic equilibrium assumptions.  Instead, the model is derived directly from Langevin dynamics, allowing it to capture both equilibrium and non-equilibrium regimes within a single formulation (Table \ref{tab:estimation_comparison}). We show through simulations that this method provides more generalizable performance in both equilibrium and non-equilibrium regimes compared to existing methods. We further demonstrate the applicability of the model in real-world settings by performing hypothesis testing on EGG recordings and analyzing the connectivity structure inferred by the model on intracortical recordings.

The key assumption that we make in our formulation is the small angle approximation in (\ref{eqn:loglikelihood:approximation:continoustime}), and the performance of our model is governed by the ratio of the sampling interval $\Delta$ to the oscillators’ intrinsic frequency $\underline{\omega}$. However, modern physiological data acquisition systems typically sample at kilohertz rates, making this assumption justifiable. A limitation of this approach is that our model takes a single deterministic value of $\omega_i$ (Table \ref{tab:estimation_comparison}). Neuromuscular cells in the stomach are unique in that they have a stereotypical oscillatory pattern in a narrow frequency band of $0.03\text{ -- }0.07\,\text{Hz}$. However, brain activity can exhibit heterogeneous, broadband dynamics, and several studies have shown that neural activity appears to coordinate traveling-wave dynamics across a wide range of frequencies \cite{gonzales2025touch}, making it difficult to define a single point $\omega_i$. In these settings, approaches assuming an equilibrium state are more appropriate as they only take phase differences across the oscillators into account \cite{cadieu2010phase, perley_coleman_2025_phase}. Nonetheless, many neural and physiological processes operate within a narrow frequency band, allowing the mean frequency to be used as $\omega_i$. 

There are natural extensions to the work proposed here. Notice that our simulation results are focused on cases where $\kappa_{ij}$ remains constant over time. However, there are many physiological systems where $\kappa_{ij}$ is time varying, such as constantly changing neural circuits with synaptic plasticity \cite{bittner2017behavioral}. Going forward, one can estimate time varying model parameters through state space modeling with the linear formulation proposed in (\ref{eqn:kappahat:matrixinverse}). These approaches may help track the evolution of coupling structure over time, enabling the identification of transient patterns such as traveling waves triggered by visual stimuli in the dorsal cortex \cite{aggarwal2022visual}.

\section*{Acknowledgements}
We would like to acknowledge support from
the Wu Tsai Neurosciences Institute and NIH 5U01NS131914.

\bibliographystyle{IEEEtran}

\vspace{1em}
\bibliography{Kuramoto-VM}

\end{document}